\documentclass[aps,prb,twocolumn,showpacs,floatfix,superscriptaddress]{revtex4}
\usepackage{graphicx,bm,amsmath,amssymb,natbib,url,epsfig,psfrag}

\begin{document}

\title{Magnetic edge states in graphene in nonuniform magnetic fields}
\author{Sunghun Park}
\author{H.-S. Sim}
\affiliation{Department of Physics, Korea Advanced Institute of
Science and Technology, Daejeon 305-701, Korea}

\date{\today}

\begin{abstract}
We theoretically study electronic properties of a graphene sheet
on $xy$ plane in a spatially nonuniform magnetic field,
$B = B_0 \hat{z}$ in one domain and $B = B_1 \hat{z}$ in the other
domain, in the quantum Hall regime and in the low-energy limit.
We find that the magnetic edge states of the Dirac fermions,
formed along the boundary between the two domains,
have features strongly dependent on
whether $B_0$ is parallel or antiparallel to $B_1$.
In the parallel case,
when the Zeeman spin splitting can be ignored,
the magnetic edge states originating from
the $n=0$ Landau levels of the two domains have dispersionless
energy levels, contrary to those from the $n \ne 0$ levels.
Here, $n$ is the graphene Landau-level index.
They become dispersive as the Zeeman splitting becomes finite
or as an electrostatic step potential is additionally applied.
In the antiparallel case,
the $n=0$ magnetic edge states split into electron-like and
hole-like current-carrying states.
The energy gap between the
electron-like and hole-like states
can be created by the Zeeman splitting or by the step potential.
These features are attributed to the fact that the pseudo-spin
of the magnetic edge states couples to the direction of the
magnetic field.
We propose an Aharonov-Bohm interferometry setup in a graphene ribbon
for experimental study of the magnetic edge states.
\end{abstract}



\maketitle

\section{Introduction}
\label{sec:Introduction}

Graphene, a two-dimensional (2D) honeycomb lattice of carbons,
has attracted much attention, because of its unusual electronic
properties.
In the low-energy regime, electrons near the two inequivalent
valleys, $K$ and $K'$, of its electronic structure can be described
by massless Dirac fermions\cite{Semenoff,Haldane}
and exhibit half-integer quantum Hall effects,\cite{Novoselov,Zhang}
in contrast to the quantum Hall effect of the conventional 2D electrons
formed in semiconductor heterostructures.\cite{Chakraborty}
The long phase coherence length and mean free path,
of micrometer order, measured\cite{Berger,Miao}
in high-quality graphenes indicate potential applications of
a graphene ribbon to coherent nanodevices.

On the other hand, the electron properties of the conventional 2D
systems have been investigated in the presence of spatially
nonuniform magnetic fields. The nonuniform fields can cause the
formation of charateristic current-carrying edge
states\cite{Mints,Muller} along the region of field gradient, which
correspond to the semi $\vec{\nabla} B \times \vec{B}$ drift motion.
These states have been refered\cite{Sim} as {\em magnetic edge
states} in the analogy to the edge states,\cite{Halperin}
corresponding to the $\vec{E} \times \vec{B}$ drift, along sample
boundaries. Their features have been studied
experimentally.\cite{Nogaret} The nonuniform fields can form various
magnetic structures such as magnetic
steps,\cite{Peeters1,Badalyan,Reijniers1} magnetic quantum
dots,\cite{Sim,Reijniers2,Novoselov2} magnetic rings,\cite{Kim}
magnetic superlattices,\cite{Carmona,Ye,Ibrahim} etc, and play a
role of characteristic barriers and resonators of electron
transport,\cite{Matulis,Sim2} which properties are very different
from those formed by electrostatic gate potential.

A graphene sheet may provide a good experimental system
for studying the effects of the nonuniform magnetic fields
as a nonuniform-field configuration can be effectively generated
by applying a uniform field to a curved sheet.\cite{Leadbeater}
In addition, the Dirac fermions in graphene can
have interesting properties under the nonuniform fields\cite{Martino},
which may be different from those of the magnetic edge states
in the conventional 2D systems, since
the Dirac fermions
have electron-like, zero-mode, and hole-like
Landau levels, as well as the pseudo-spins representing
the two sub-lattice sites of the honeycomb lattice.
Therefore, it may be valuable to study the electron properties of
graphene in a nonuniform magnetic field,
which is the aim of the present work.

In this theoretical work, we study the electronic
structures of a graphene sheet (on $xy$ plane)
in a spatially nonuniform magnetic field of step shape,
$\vec{B}=B_0 \hat{z}$ for $x<0$ and
$\vec{B}=B_1 \hat{z}$ for $x>0$, in the integer quantum Hall regime,
based on a noninteracting-electron approach.
By solving the Dirac equation,
we first investigate the low-energy properties of
the magnetic edge states formed along
the boundary ($x=0$) between the two domains with different
fields $B_0$ and $B_1$, when the Zeeman effect is negligible.
They are found to strongly
depend on whether $B_0$ is parallel or antiparallel to $B_1$.
In the palallel case of $\gamma \equiv B_1/B_0 > 0$,
the magnetic edge states originating from the $n=0$ Landau levels
of the two domains are dispersionless,
contrary to those from the $n \ne 0$ levels,
where $n$ is the Landau level index.
In the antiparallel case of $\gamma < 0$,
the $n=0$ magnetic edge states
split into electron-like and hole-like levels near the boundary.
These features, which are absent in the magnetic edge states
of the conventional 2D electrons, are attributed to the fact that
the pseudo-spin of the magnetic edge states
couples to the direction of the magnetic field.
On the other hand, the features of
the $n \ne 0 $ magnetic edge states
are similar to those of the conventional cases.

We further study the magnetic edge states
in the presence of an additional electrostatic step potential,
$V(x) = V_0$ for $x < 0$ and $V(x) = V_1$ for $x > 0$.
For $\gamma > 0$, the $n=0$ magnetic edge states become dispersive,
while for $\gamma < 0$
an energy gap becomes created around the bipolar region
in the spectrum of
the magnetic edge states.
Similar features can be found when the Zeeman spin splitting is finite,
because in the nonuniform field of step shape
the Zeeman effect behaves as the step potential.

Finally, we suggest an interferometry setup formed in a graphene
ribbon for experimental study.
In this setup, the magnetic edge states can provide partial paths
of a full Aharonov-Bohm interference loop, therefore
the properties of the magnetic edge states such as the gap of
their energy spectra can be investigated.
We numerically calculate the transmission probability through the setup
by using the tight-binding method
and the Green function technique.\cite{Meir,Datta,Sim3}
The results are consistent with the features of the
magnetic edge states obtained by solving the Dirac equation.
We also derive the transmission probability, based on the scattering
matrix formalism, and use it to analyze the numerical results.

This paper is organized as follows.
The magnetic edge states are studied without and with
the electrostatic step potential
in Secs.~\ref{sec:magneticedge} and \ref{sec:step}, respectively.
The Zeeman spin splitting is considered in Sec.~\ref{sec:spin},
and the graphene interferometry is
suggested and investigated in Sec.~\ref{sec:interfero}.
In Sec.~\ref{sec:summary}, this work is summarized.
Throughout this work,
we ignore the intervalley mixing due to the nonuniform field,
the validity of which is discussed in Appendix~\ref{APP1}.
In Appendix~\ref{APP2} we derive
the transmission probability,
which may be applicable to other graphene interferometry setups
with slight modification.

\section{Magnetic edge states}
\label{sec:magneticedge}

We consider a graphene sheet (on $xy$ plane)
in the nonuniform magnetic field of the step configuration,
\begin{eqnarray} \label{NONUNIFORMB}
    \vec{B}(x) = \left\{
      \begin{array}{ll}
        B_{0} \hat{z}, \,\,\, & \hbox{$x<0$} \\
        B_{1} \hat{z}, \,\,\, & \hbox{$x>0$}
      \end{array}
    \right.
\end{eqnarray}
in the integer quantum Hall regime.
Without loss of generality, $B_0$ is chosen to be positive,
and $\gamma \equiv B_1 / B_0$ is either positive or negative.
This configuration may be
realized with field gradient $dB/dx$ less than
$10^{4} \,\,\, \textrm{T} \cdot \textrm{{\AA}}^{-1}$
and with not-too-strong field strengths
$B_0$ and $|B_1|$ (less than, e.g., 100 T).
In this case, the mixing between the $K$ and $K'$ valleys
due to the nonuniform field
can be ignored (see Appendix A),
and the electrons in each valley
can be seperately described by the Dirac equation,
\begin{eqnarray}
[ v \vec{\sigma}_J \cdot \vec{\Pi} + V (x) ] \Psi_J
& = & E \Psi_J,
\label{DIRAC}
\end{eqnarray}
in the low-energy approximation.
Here, $J \in \{ K, K' \}$ is the valley index,
$v = \sqrt{3} a t / (2 \hbar) \sim 10^6 \, \, \textrm{m}
\cdot \textrm{s}^{-1}$, $a$ is the lattice constant of graphene,
$t$ is the hopping energy between two nearest neighbor sites,
$\vec{\sigma}_K = (\sigma_x,\sigma_y)$ and
$\vec{\sigma}_{K'} = (-\sigma_x, \sigma_y)$
are constructed by the Pauli matrices,
$\vec{\Pi} \equiv \vec{p} + e \vec{A}$,
$\vec{p}$ is the momentum measured relative to the valley center
($K$ or $K'$ point), $\vec{A}$ is the vector potential,
and
$e (> 0)$ is the electron charge.
The electrostatic potential $V(x)$ applied to the sheet
must be a smoothly varying function of $x$.
The detailed form of $V(x)$ will be specified in Sec. \ref{sec:step}.
The components of the pseudo-spinor $\Psi_J$ represent the wave functions
of the two sub-lattice sites (denoted by $A$ and $B$ hereafter)
of a unit cell of graphene.
The Dirac equations for the $K$ and $K'$ valleys
are connected to each other by a unitary
transformation $U = \sigma_y$.
This feature leads that
the two equations have the same energy levels
and that their wave functions have the relation, $\Psi_{K'} = U \Psi_K$.
Therefore, it is enough to solve the Dirac eqaution for the $K$ valley
only.

Before studying the nonuniform-field cases, we briefly discuss
the case of a uniform magnetic field with strength $B$.
In this case,
the Landau levels are found\cite{Novoselov} to be
\begin{eqnarray}
E_n = \textrm{Sgn}(n) v \sqrt{2 |n| \hbar e B },
\end{eqnarray}
where $n = 0, \pm 1, \pm2, \cdots$ is the Landau level index.
The levels with $n>0$ are often refered as electron-like,
while those with $n<0$ as hole-like.
In fact, one can obtain the Landau levels from the square
of the Dirac Hamiltonian,
\begin{eqnarray}
E_n^2 = 2 \hbar e v^2 B (m + 1/2 \pm 1/2)
\label{ESCQURE}
\end{eqnarray}
for the $K$ valley, with
$|n| = m + 1/2 \pm 1/2$ and
$m = 0,1,2,\cdots$.
For later discussion, it is worthwhile to analyze $E_n^2$.
The harmonic term with $B(m + 1/2)$ is nothing but
the Landau level of the conventional 2D electrons,
while the next term $\pm B/2$ can be interpreted as
the effective Zeeman effect of the pseudo-spin;
$B/2$ for pseudo-spin up and $-B/2$ for down.
Each Landau level with $n \ne 0$ is composed of the pseudo-spin
up (with $m = |n| - 1$) and pseudo-spin down (with $m = |n|$)
states, while the $n=0$ level comes only from the pseudo-spin
down (up) states with $m=0$ in the $K$ ($K'$) valley.
The $n=0$ level is independent of
$B$ as the harmonic and effective Zeeman terms
exactly cancel each other.

Turning back to the nonuniform field in Eq. (\ref{NONUNIFORMB}),
we consider the case without electrostatic potential,
$V(x) = 0$, in this section.
We choose the vector potential as
$\vec{A}(x) = B_{0} x \hat{y}$ for $x<0$ and
$\vec{A}(x) = B_{1} x \hat{y}$ for $x>0$.
This choice of $\vec{A}$ is useful, as
the solution of Eq. (\ref{DIRAC}) for the $K$ valley
can be written as
$\Psi_K^\dagger (x) = e^{-iky} \Phi^\dagger (x)
= e^{-iky} (\phi^{\ast}_{A}(x),\phi^{\ast}_{B}(x))$,
where $\phi_{j}$ is the wave function of the sub-lattice
site $j \in \{A,B \}$ and $\hbar k$ is the eigenvalue of $p_{y}$.
Hereafter we will measure
energy and length in units of $E_1 (\equiv \sqrt{2\hbar
v^{2}eB_{0}})$ and $l_{B}(\equiv \sqrt{\hbar/(eB_{0})})$,
where $l_B$ and $E_1$ are the magnetic length and
the energy gap between the $n=0$ and $n=1$ Landau levels, respectively,
of the bulk region with $B_0$;
in these units, the Landau levels are
$\sqrt{n}$ at $x \ll - 1$ while
$\textrm{Sgn}(n) \sqrt{|\gamma| n}$ at $x \gg 1 / \sqrt{|\gamma|}$.
Then the equation for $\phi_{j = A,B}$ is found to be
\begin{eqnarray}
- \frac{d^2 \phi_{j,k} (x)}{dx^2}
+ 2[V^j_{\rm eff}(x,k) - E^2_{n,k}] \phi_{j,k}(x) = 0.
\label{DIRAC_SUB}
\end{eqnarray}
The effective potential\cite{Sim} $V_{\rm eff}^j$ is harmonic,
\begin{equation}
V^j_{\rm eff}(x,k)=\left\{
  \begin{array}{ll}
    \frac{1}{2}(x+k)^2 + s_j \frac{1}{2}, & \hbox{$x<0$} \\ \\
    \frac{1}{2}(\gamma x + k)^2 + s_j \frac{\gamma}{2}, & \hbox{$x>0,$}
  \end{array}
\right.
\label{EFFV}
\end{equation}
where $s_{j = A} = 1$ and $s_{j=B} = -1$.
The equation (\ref{DIRAC_SUB}) has the form of the usual
Schr\"{o}dinger equation
with potential $V_{\rm eff}^j$ and eigenvalue $E^2_{n,k}$.
Therefore, $V_{\rm eff}^j$ is useful for understanding
$\phi_{j,k} (x)$ and $E_{n,k}$.
The form of $V_{\rm eff}^j$ in Eq. (\ref{EFFV}) shows that
the pseudo-spin ($s_j$) couples to the {\em direction}
($\gamma$) of the magnetic field.

\begin{figure}[tb]
\includegraphics[width=0.45\textwidth]{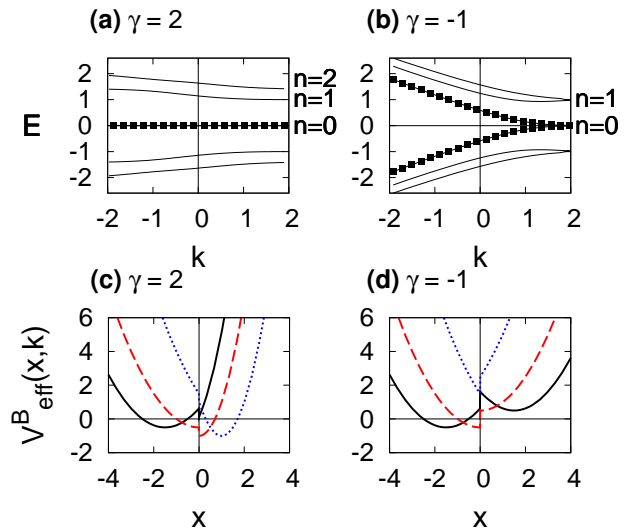}
\caption{
(Color online)
Upper panels:
Energy spectra $E_{n,k}$ for (a) $\gamma=2$ and (b) $\gamma=-1$.
The energy levels of the $n=0$ magnetic edge states
are highlighted by the filled squares.
For $\gamma > 0$, the $n=0$ levels are dispersionless while
they split into electron-like and hole-like levels for $\gamma < 0$.
Lower panels:
Effective potential $V^{j=B}_{\rm eff}(x,k)$ in Eq.(\ref{EFFV})
for (c) $\gamma=2$ and (d) $\gamma=-1$.
Different values of $k=1.5$ (solid line), 0 (dashed), -2 (dotted)
are chosen.
Energy and length are measured in units of
$E_1 (\equiv \sqrt{2\hbar v^{2}eB_{0}})$
and $l_{B}(\equiv \sqrt{\hbar/(eB_{0})})$, respectively.
}
\label{fig1:energy}
\end{figure}

The solution $\phi_{j,k}^{<}$ of Eq. (\ref{DIRAC_SUB}) for $x<0$
can be expressed in terms of
the parabolic cylinder functions\cite{Lebedev} $D_{\nu}(z)$ as
\begin{equation}\label{SOL_NEG}
    \left(
    \begin{array}{c}
      \phi^<_{A,k} \\
      \phi^<_{B,k}
    \end{array}
    \right)
    \propto
    \left(
    \begin{array}{c}
      i E D_{E^2 - 1}(-\sqrt{2}(x+k)) \\
      D_{E^2}(-\sqrt{2}(x+k))
    \end{array}
    \right).
\end{equation}
On the other hand, for $x > 0$, the solution $\phi_{j,k}^>$
is found to be dependent on the sign of $\gamma$, i.e., on
whether $B_1$ is either parallel or antiparallel to $B_0$, as
for $\gamma > 0$
\begin{equation}\label{SOL_POS_1}
    \left(
    \begin{array}{c}
      \phi^>_{A,k} \\
      \phi^>_{B,k}
    \end{array}
    \right)
    \propto
    \left(
    \begin{array}{c}
      - i \frac{E}{\sqrt{\gamma}} D_{\frac{E^2}{\gamma} - 1}
(\sqrt{2 \gamma}(x + \frac{k}{\gamma})) \\
      D_{\frac{E^2}{\gamma}}(\sqrt{2 \gamma}(x+\frac{k}{\gamma}))
    \end{array}
    \right),
\end{equation}
and for $\gamma < 0$
\begin{equation}\label{SOL_POS_2}
    \left(
    \begin{array}{c}
      \phi^>_{A,k} \\
      \phi^>_{B,k}
    \end{array}
    \right)
    \propto
    \left(
    \begin{array}{c}
      D_{\frac{E^2}{|\gamma|}}(\sqrt{2 |\gamma|}(x+\frac{k}{\gamma})) \\
      - i \frac{E}{\sqrt{|\gamma|}} D_{\frac{E^2}{|\gamma|} - 1}
(\sqrt{2 |\gamma|}(x + \frac{k}{\gamma}))
    \end{array}
    \right).
\end{equation}
The energy eigenvalue
$E_{n,k}$ can be obtained under the boundary condition of
the continuity of
the wave functions in Eqs.~(\ref{SOL_NEG})-(\ref{SOL_POS_2})
at $x=0$.
For the two cases of $\gamma = 2$ and $\gamma = -1$,
$E_{n,k}$ is drawn in Fig.~$\ref{fig1:energy}$.
These two specific cases of $\gamma$
are enough to understand the characteristics of the magnetic edge states.

The energy levels strongly depend
on whether $B_1$ is parallel or antiparallel to $B_0$.
In the parallel case of $\gamma > 0$,
for each $n$ the energy levels gradually change from
$\textrm{Sgn}(n) \sqrt{|n|}$ to $\textrm{Sgn}(n) \sqrt{\gamma |n|}$
as $k$ decreases from positive to negative values
[see the $\gamma = 2$ case in Fig.~\ref{fig1:energy}(a)].
This feature can be understood from the effective potential
[see Eq.~(\ref{EFFV}) and Fig.~\ref{fig1:energy}(c)].
As $k$ increases from negative to positive values, the bottom
of $V_{\rm eff}^j(x)$ moves from the region with $B_1$ to that
with $B_0$, passing the boundary $x=0$ around $k=0$.
Therefore, for $k \ll 0$,
the eigenstates have the Landau levels
$\textrm{Sgn}(n) \sqrt{\gamma |n|}$ and
localize around $x = - k/\gamma$
(in the unit of $\l_B$), while
for $k \gg 0$, they have
$\textrm{Sgn}(n) \sqrt{|n|}$  and localize around $x = - k$.
Around $k = 0$, the two levels of
$\textrm{Sgn}(n) \sqrt{\gamma |n|}$
and
$\textrm{Sgn}(n) \sqrt{|n|}$
connect smoothly
and the corresponding states are localized around $x=0$.
These states have been called\cite{Sim} magnetic edge states
and they can carry current along the boundary $x=0$.

Contrary to the corresponding states in the conventional 2D systems,
the magnetic edge states in graphene
have the following different features for $\gamma > 1$.
First, the edge states with $n > 0$ behave as electrons,
with dispersion $dE_n / dk < 0$ while those with $n < 0$ as holes
with $dE_n / dk > 0$.
Second, the edge states with $n = 0$ are dispersionless and
carry no current, {\em regardless} of $\gamma (> 0)$.
These features come from the nature of the Dirac fermions.
Especially, the second one can be understood
from the fact that for $\gamma > 0$,
the effective Zeeman and harmonic contributions
to $E^2_{n=0}$ cancel each other in the both sides of $x=0$,
as discussed around Eq. (\ref{ESCQURE})
and as shown in the term of $V_{\rm eff}^j$
representing the coupling of pseudo-spin to field direction
[see Eq. (\ref{EFFV})].
The case of $0 < \gamma < 1$ can be understood in a similar way.

Next, we discuss the case of $\gamma < 0$, which is very different
from the $\gamma > 0$ case [see Fig.~\ref{fig1:energy}(b)]. For
large positive $k$, the eigenstates have the Landau levels of
$\textrm{Sgn}(n) \sqrt{|n|}$ or $\textrm{Sgn}(n) \sqrt{|\gamma n|}$,
while for large negative $k$ $(\ll 0)$, their energy either
increases (showing electron-like behavior) or decreases (hole-like).
This feature can be understood from $V^j_{\rm eff}(x,k)$. As shown
in Fig.~\ref{fig1:energy}(d), for large positive $k$, the two local
harmonic wells of $V^j_{\rm eff}(x,k)$ occur far from the boundary
of $x = 0$, resulting in the Landau levels localized in each well.
As $k$ decreases to negative values, the two local wells move toward
$x=0$ and become to merge into a single well (not harmonic anymore)
at $x=0$, and then the bottom of the merged well increases.
Therefore, the eigenstates with $k \lesssim 0$ are magnetic edge
states localized at $x=0$ and they can carry current along $x=0$.
They are either electron-like ($dE_{n \ge 0}/dk < 0$) or hole-like
($dE_{n \le 0}/dk > 0$). One can estimate their group velocity for
$k \ll 0$ as $(1/\hbar) d E_n / d k \sim \pm v$ from the minimum
value of the merged well of $V_{\rm eff}^B$, where $+ (-)$ stands for
the hole(electron)-like states. In this case of $\gamma < 0$, where
the magnetic fields $B_0$ and $B_1$ are antiparallel, the
eigenstates in $n\neq0$ Landau levels correspond to classical
motions, so called snake orbits\cite{Mints,Muller,Sim,Badalyan},
while those in the $n=0$ level have no clear correspondence to
classical motions as they have both electron and hole characters.

For $\gamma < 0$,
the effective Zeeman contribution to $E_n^2$
has the opposite sign between the domains of $x > 0$ and $x < 0$,
as shown in Eq. (\ref{EFFV}).
This coupling of pseudo-spin to field direction
causes an energy barrier at $x = 0$;
for example, a pseudo-spin down state
has larger effective Zeeman
contribution at $x > 0$ than at $x < 0$.
As a result, as $k$ decreases, the magnetic edge state
becomes more confined around $x=0$ due to the effective-Zeeman barrier.
This pseudo-spin feature, which enhances the splitting
into electron-like and hole-like states,
is absent in the magnetic edge states of the conventional 2D electrons.

\section{Magnetic edge states in an electrostatic step potential}
\label{sec:step}

In this section, we consider an additional electrostatic potential
$V(x)$
of step shape,
\begin{equation} \label{stepV}
    V(x)=\left\{
      \begin{array}{ll}
        V_{0}, \,\,\, & \hbox{$x<0$} \\
        V_{1}, \,\,\, & \hbox{$x>0$},
      \end{array}
    \right.
\end{equation}
to the nonuniform magnetic field in Eq. (\ref{NONUNIFORMB}).
Here $V_0$ and $V_1$ are constant.
This potential gives rise to characteristic modification
of the $n=0$ magnetic edge states, such as the creation of
energy gap for $\gamma < 0$,
as will be seen below.
Moreover, the modification is directly applicable to
the case where the Zeeman spin splitting is finite, as will be
studied in Sec.~\ref{sec:spin}.

\begin{figure}[tb]
\includegraphics[width=0.41\textwidth]{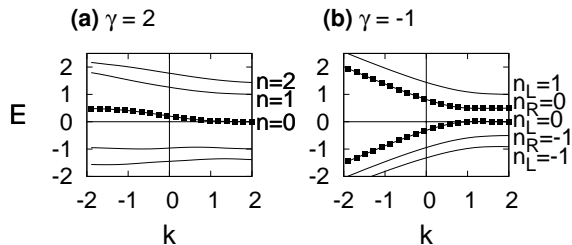}
\caption{
The same as in Fig.~\ref{fig1:energy} but
in the presence of the step potential $V(x)$
with $V_0 = 0$ and $V_1 = 0.5$.
For $\gamma = -1$,
the energy levels are labeled by $n_L$ and $n_R$,
which are the indices of the Landau levels localized
at $x \ll - 1$ and at $x \gg 1/\sqrt{|\gamma|}$, respectively (see text).
The $n=0$ levels become dispersive for $\gamma > 0$ while
the energy gap opens between electron-like ($n_R = 0$)
and hole-like levels ($n_L = 0$) for $\gamma < 0$.
}
\label{fig2:energy2}
\end{figure}

The step potential $V(x)$ is assumed to be smoothly varying
in the length scale of the lattice constant $a$. Then,
we can still ignore the intervalley mixing and solve the Dirac
equation in Eq. (\ref{DIRAC}) in the same way as above.
In Fig.~\ref{fig2:energy2}, choosing $V_{0}=0$ and $V_{1}=0.5$
(in the units of $E_1$ and $l_B$),
we draw the energy spectra of the magnetic edge states
for $\gamma=2$ and $\gamma=-1$.

The features of the energy spectra are discussed below.
For $\gamma > 0$,
the eigenstates with large positive (negative) $k$
have the Landau levels shifted by $V_0$ ($V_1$) and do not carry
current, while the magnetic states around $k=0$ have the energy
smoothly connecting the Landau levels of the large positive
and negative $k$'s.
We point out that the $n=0$ magnetic edge states carry current
due to the potential step $V(x)$.
On the other hand, for $\gamma < 0$,
the eigenstates with large positive $k$ have the Landau levels
shifted by $V_0$ or $V_1$, depending on whether they are localized
in the bulk region of $x < 0$ or $x > 0$.
For convenience, we introduce the Landau-level indices $n_L$
and $n_R$ for those localized in $x < 0$ and in $x > 0$, respectively.
In this case,
the energy levels with $n=0$ split, opening the {\em energy gap}
between the electron-like and hole-like magnetic edge states,
in contrary to the case without the step potential.
For $\gamma = -1$ and $V_1 - V_0 = 0.5$,
the gap size is the same
as the step height $V_1 - V_0$.

\begin{figure}[tp]
\includegraphics[width=0.32\textwidth]{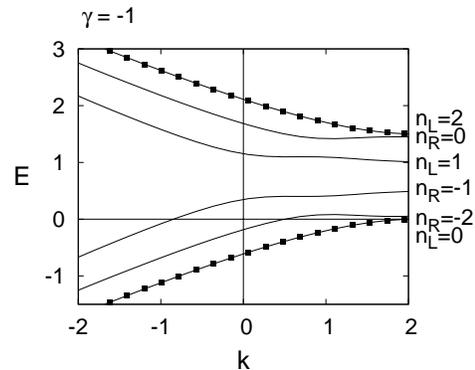}
  \caption{
The same as in Fig.~\ref{fig2:energy2}(b) but
with $(V_0,V_1) = (0,1.5)$.
}\label{fig3}
\end{figure}

We further study the energy gap
with varying the height $V_1 - V_0$
for $\gamma = -1$.
In Fig.~\ref{fig3},
we choose $V_1 - V_0 = 1.5$,
which is larger than the energy spacing
between the $n=0$ and $n=1$ Landau levels,
in contrary to the case of Fig.~\ref{fig2:energy2}(b) where
the height (= 0.5) is smaller.
In this case, the energy gap of the
magnetic edge states occurs between the $n_L = 1$ and $n_R = -1$
Landau levels. Moreover, the gap size is no longer
the same as the step height, but corresponds to the energy
difference (= 0.5) between the $n_L = 1$ and $n_R = -1$ levels.

The above feature can be understood as follows.
For $\gamma = -1$, the effective Hamiltonian in Eq. (\ref{DIRAC})
is odd under the inversion operator $\mathcal{R}_{\rm inv}$,
$\mathcal{R}_{\rm inv} \vec{x} = -\vec{x}$,
\begin{eqnarray}
& & \mathcal{R}_{\rm inv}
\big[v \vec{\sigma}_J \cdot \vec{\Pi}
+ V(x) - \frac{V_0 + V_1}{2} \big] \mathcal{R}_{\rm inv}
\mathcal{R}_{\rm inv} \Psi_J  \nonumber \\
& = & - \big[v \vec{\sigma}_J \cdot \vec{\Pi}
+ V(x) - \frac{V_0+ V_1}{2} \big] \mathcal{R}_{\rm inv}
\Psi_J  \nonumber \\
& = &  (E - \frac{V_0+ V_1}{2}) \mathcal{R}_{\rm inv} \Psi_J.
\label{DIRAC2}
\end{eqnarray}
This property is consistent with the facts that the gap center
is located at $E_c \equiv (V_0 + V_1)/2$
and that the gap size is determined by the energy difference
between the Landau levels just below and above $E_c$.
Similarly to the case in Fig.~\ref{fig1:energy},
it can be also understood\cite{EFFECTIVE} from $V_{\rm eff}^j$
that the states with $E > E_c$ are electron-like while those with
$E < E_c$ are hole-like.

\section{Zeeman splitting of the magnetic edge states}
\label{sec:spin}

So far, we have ignored the spin degree of freedom.
Recently, Abanin, Lee, and Levitov discussed\cite{Abanin}
that in graphene
the Zeeman splitting can be smaller than the Landau
energy gap only by the factor of about $10^{-1}$,
due to the exchange interaction, and that
it can play an important role in the edge-channel transport
in the quantum Hall regime.
In this section, we discuss the effect of the Zeeman splitting
on the magnetic edge states.

\begin{figure}
\includegraphics[width=0.43\textwidth]{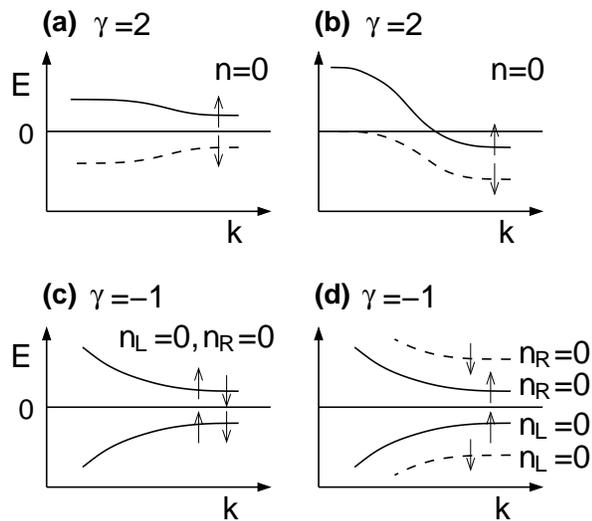}
\caption{
Schematic spin-dependent
energy dispersion of the $n=0$ magnetic edge states
in the presence of the Zeeman splitting, 2 $|V_s^{\rm Z}| = 0.2$.
We choose the parameters $(\gamma,V_0,V_1)$ as
(a) (2,0,0), (b) (2,-0.2,0.2), (c) (-1,0,0), (d) (-1,-0.2,0.2).
}\label{fig4}
\end{figure}

In the nonuniform field $B(x)$ in Eq. (\ref{NONUNIFORMB}),
the Zeeman splitting
behaves as a spin-dependent step potential,
\begin{equation}\label{zp}
    V^{\rm Z}_s (x) = s \Delta_Z B(x),
\end{equation}
where $\Delta_{\rm Z} = g^* \mu_B / E_1$,
$s = 1/2$ ($s = -1/2$) for spin-up (down) electrons,
and $\mu_B$ is the Bohr magneton.
The exchange enhancement\cite{Abanin}
of the $g$-factor can be taken into account in $g^*$.
We assume that $V^{\rm Z}_s \ll 1$ for convenience.

The Zeeman splitting can be considered as the spin-dependent shift
of the step potential, $V(x) \to V(x) + V^{\rm Z}_s (x)$.
The resulting magnetic edge states can be easily understand
from the features discussed in Sec.~\ref{sec:step}.
In Fig.~\ref{fig4}, we draw schematic energy dispersions of the
$n=0$ magnetic edge states; the extension to $n \ne 0$ states is trivial.
For $\gamma > 0$, the $n=0$ states are dispersive near $x = 0$, as
shown in Figs.~\ref{fig4}(a) and (b).
The energy difference between the spin-up and down states
varies from $\gamma \Delta_{\rm Z}$ to $\Delta_{\rm Z}$ as $k$ increases.
And, the average value of the spin-up and down energy levels is shifted
by $V_0$ ($V_1$) at large positive (negative) $k$.
On the other hand, for $\gamma < 0$,
the energy gap of the $n=0$ magnetic edge states exists even
without the electrostatic step potential [see Fig.~\ref{fig4}(c)].
For both the cases of $\gamma > 0$ and $\gamma < 0$,
the sign of the drift velocity ($  \sim dE/dk$) is either
positive or negative, depending on $\Delta_{\rm Z}$, $V_0$, and $V_1$.
These spin-split dispersions show that
spin-polarized current can emerge between two magnetic domains.

\section{Graphene interferometry}
\label{sec:interfero}

\begin{figure}
\includegraphics[width=0.45\textwidth]{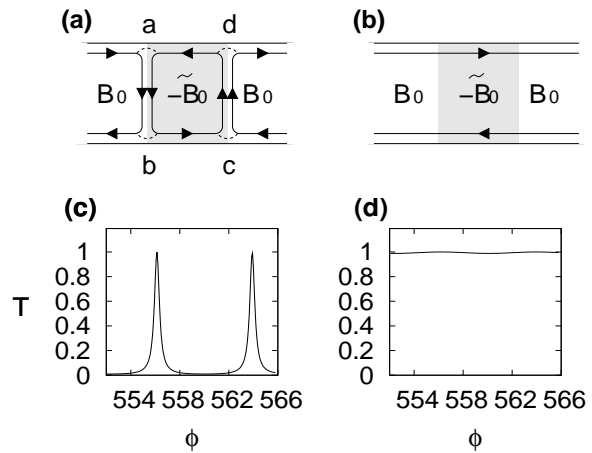}
\caption{
Upper panel: Schematic diagram of a graphene-ribbon interferometry,
which consists of the left (with magnetic field $B_0$),
middle (with $- \tilde{B}_0$), and right (with $B_0$) regions.
There is one edge channel along the ribbon edges,
while
there can be magnetic edge states along
the left and right boundaries,
$\bar{ab}$ and $\bar{cd}$, of the middle region,
depending on the energy of the states and the field configuration
of ($B_0$, $-\tilde{B}_0$).
In (a),
there appear two magnetic edge states along each boundary,
and thus the Aharonov-Bohm interference,
while no state and no interference exists in (b).
Lower panel: Aharonov-Bohm effect in the transmission $T$
through the interferometry,
as a function of the Aharonov-Bohm flux $\phi$.
The case of (a) is plotted in (c), while that of (b) in (d).
}\label{fig5}
\end{figure}

In this section, we propose an interferometry setup for
studying magnetic edge states in a graphene ribbon.
We focus on the case of $\gamma = -1$ in Fig.~\ref{fig2:energy2}(b)
and demonstrate that
the energy gap of the magnetic edge states
can be directly studied by observing the Aharonov-Bohm
interference of the setup.
The interferometry setup is useful as well for the other cases of $\gamma$.

We consider a ribbon with armchair edge; a setup with zigzag edge
will show a similar result. The ribbon consists of three parts, the
current source in the left, the middle scattering region, and the
drain in the right. A nonuniform magnetic field is applied such that
$B = B_0 \hat{z}$ in the source and drain while $B = \tilde{B}_0
\hat{z} \simeq - B_0 \hat{z}$ in the middle region (see
Fig.~\ref{fig5}). At the same time, a constant electrostatic
potential $V_0 = 0$ is applied in the source and drain while $V_1 =
0.5$ in the middle region. Then, the magnetic edge states form along
the left and right boundaries $\bar{ab}$ and $\bar{cd}$ of the
middle region, while along each boundary of the ribbon, there is one
edge channel, which is formed as a mixture of the contribution of
the $K$ and $K'$ valleys.\cite{Brey} The magnetic edge states are
the same as those studied in Fig.~\ref{fig2:energy2}(b), when their
position separates from the ribbon edge by more than the scale of
magnetic length, $l_B \equiv \sqrt{\hbar / (eB_0)}$, so that the
overlap between the magnetic edge states and the ribbon-edge
channels is negligible. Note that at each of the scattering points
$a$-$d$ in Fig.~\ref{fig5}(a), the number of the incoming channels
is the same as that of the outgoing channels, so that the current is
conserved.

The formation of the magnetic edge states depends on energy.
In the energy range $\in [0.5, 1]$ of Fig.~\ref{fig2:energy2}(b),
where each valley supports only one magnetic edge channel,
one has the edge-channel transport shown in Fig.~\ref{fig5}(a).
In this case, there appears an Aharonov-Bohm loop around the middle region,
which is supported by the two counterpropagating edge channels
along the upper and lower ribbon edges and by the magnetic edge states
along the boundaries $\bar{ab}$ and $\bar{cd}$. As a function of
$\tilde{B}_0$, one can observe the Aharonov-Bohm interference in
the transmission through the setup.
On the other hand, in the energy range $\in [0, 0.5]$, where
there is no magnetic edge state along $\bar{ab}$ and $\bar{cd}$
[see Fig.~\ref{fig5}(b)],
no Aharonov-Bohm interference can be observed.
Therefore, the Aharonov-Bohm loop can be formed, depending on
whether the magnetic edge channels exist along the boundaries
$\bar{ab}$ and $\bar{cd}$.
This property allows one to measure the gap of the magnetic
edge states by varying the energy of the incoming edge channel
and by modulating $\tilde{B}_0$.

We confirm the above proposal numerically by calculating the
transmission probability $T(E)$ through the setup, using the
tight-binding method\cite{Brey2} and the Green's function
approach.\cite{Meir,Datta} Here we skip the details of the method
and instead refer Refs. \cite{Sim2,Sim3}. The effect of the magnetic
field is taken into account by the Peierls phase. The strength of
$B_0$ is set to be about 800~T. At the boundaries $\bar{ab}$ and
$\bar{cd}$, the magnetic field spatially varies linearly from $B_0$
to $\tilde{B}_0$ over the length scales of $3 l_B$. We choose the
ribbon width $|\bar{ab}|$ and the width $|\bar{bc}|$ of the middle
region as about $15 l_B$ and $30 l_B$, respectively, so that we can
ignore the overlap between the edge channels propagating oppositely
to each other.

In Fig.~\ref{fig5},
we plot $T(E)$
for $E = 0.65$ and $E = 0.35$
as a function of the Aharonov-Bohm flux
$\phi \equiv 2\pi |\tilde{B}_{0}|S/\phi_{0}$, where $E$ is the energy
of the incoming states to the setup, $S$ is the area of the
middle region,
and $\phi_{0}=h/e$ is the flux quantum.
As expected, for $E = 0.65$, one has the interference,
while not for $E = 0.35$.
This confirms the proposal discussed above.

Finally, we briefly analyze the numerical result for $E = 0.65$
in Fig.~\ref{fig5}(a).
The period of the Aharonov-Bohm oscillation is found to be
$\Delta \phi = 7.8 =  1.24 \times 2 \pi$.
The fact that $\Delta \phi / (2 \pi)$ is larger than 1
indicates that the actual Aharonov-Bohm loop
has smaller area than $S$, which is reasonable.
The lineshape of the interference can be
analyzed by using the expression of $T$
in Eq. (\ref{tran}) derived in Appendix \ref{APP2}.
The lineshape is well fitted
by Eq. (\ref{tran}) with parameters of $\alpha = \beta = 0.5$
and $t_a = t_b = t_c = t_d = 0.23$. From this fitting,
one can get the information of the scattering between
the edge channels along the ribbon edges and the magnetic edge
channels along $\bar{ab}$ and $\bar{cd}$.

\section{Summary}
\label{sec:summary}

We have studied the magnetic edge states formed along the boundary
between the two domains with different magnetic fields $B_0$
and $B_1$ in graphene.
It turns out that
the magnetic edge states have very different features from
those of the conventional 2D electrons, since the formers
have pseudo-spin which couples to the direction of the magnetic
field.
As a result, the $n=0$ magnetic edge states are dispersionless
for $\gamma \equiv B_1 / B_0 > 0$ while they split into
electron-like and hole-like current carrying states for $\gamma < 0$.
The Zeeman spin splitting or the additional electrostatic step potential
can make the $n=0$ states dispersive for $\gamma > 0$
and open energy gap in the bipolar region for $\gamma < 0$.
These features show interesting manifestation of the Dirac fermions in
graphene, and
the magnetic edge states
can play a special role in the transport of the Dirac fermions
in a nonuniform magnetic field,
such as spin-polarized current along the boundaries of magnetic domains.

We are supported by Korean Research Foundation Grants
(KRF-2005-084-C00007,KRF-2006-331-C00118).

{\em Note added:} During the preparation of this manuscript, we have
been aware of two preprints\cite{Rakyta,Ghosh} where the energy
dispersion and current density of the snake states in a nonuniform
magnetic field of waveguide shape are studied. Their results
partially overlap with our results for the case of $\gamma < 0$ in
the Section~\ref{sec:magneticedge}.

\appendix

\section{Inter-valley scattering in nonuniform magnetic fields}
\label{APP1}

In Appendix \ref{APP1}, based on the tight-binding method,
we show that the mixing between the $K$ and $K'$ valleys
due to a spatially nonuniform magnetic field can be ignored when
the field strength and the gradient of the field are much smaller
than $10^4$~T and $10^4 \,\, \textrm{T} \cdot \textrm{\AA}^{-1}$,
respectively,
which is achieved in current experimental studies.

We first discuss the matrix elements of the tight-binding Hamiltonian
of graphene
in the presence of an external magnetic field.
For each sub-lattice site $j = A,B$,
the Bloch wave functions of an electron is written as
\begin{eqnarray}
\Phi^j_{\vec{k}} (\vec{r}) =
\frac{1}{\sqrt{N}} \sum_{\vec{R}_j}
e^{i \vec{k} \cdot \vec{R}_j - i \frac{e}{\hbar} \int^{\vec{r}}_{\vec{R}_j}
\vec{A} \cdot d \vec{r}_1} \phi(\vec{r} - \vec{R}_j),
\end{eqnarray}
where
the sum runs over
the potisions $\vec{R}_j$ of site $j$,
$N$ is the number of unit cells,
$\vec{A} (\vec{r}_1)$ is the vector potential,
$\vec{k}$ is the momentum of the states, and
$\phi (\vec{r})$ is the wave function of electrons participating
in the $\pi$ bonding.
The matrix element
$\langle \Phi^{B}_{\vec{k}'} | H | \Phi^{A}_{\vec{k}} \rangle$
of the tight-binding Hamiltonian
for the nearest-neighbor hopping is found to be
\begin{eqnarray}
\langle \Phi^{B}_{\vec{k}'} | H | \Phi^{A}_{\vec{k}} \rangle
& = & \frac{t}{N} {\sum}'
e^{i\frac{e}{\hbar} \int_\mathbf{C} \vec{A} \cdot d \vec{r}_1}
e^{i \vec{k} \cdot \vec{R}_{A} - i \vec{k}' \cdot \vec{R}_{B}},
\label{TB_Hamiltonian}
\end{eqnarray}
where $t$ is the hopping energy between two nearest neighbor sites
in the tight-binding scheme,
the sum ${\sum}'$ runs over the nearest-neighbor site pairs
of $A$ and $B$,
and $\mathbf{C}$ denotes the path connecting the site pair.

Using the matrix element in Eq. (\ref{TB_Hamiltonian}),
one can estimate the effect of the magnetic field on the intervalley
mixing. To do so, we assume
that $\vec{k}$ and $\vec{k}'$ are located near the $K$ and $K'$ points,
respectively.
When a uniform magnetic field is applied, the
path integration can be estimated, in terms of
the magnetic length $l_B$ and the lattice constant $a$,
as
$\frac{e}{\hbar} \int_{\mathbf{C}} \vec{A}(\vec{r}_1) \cdot d\vec{r}_1
\sim a^{2}/l^{2}_{B}$.
For
\begin{eqnarray}
a^2/l_B^2 \ll |\vec{K} - \vec{K}'| a,
\label{NOMIX}
\end{eqnarray}
the intervalley mixing is negligible, since
$\langle \Phi^{B}_{\vec{k}'} | H | \Phi^{A}_{\vec{k}} \rangle
 \simeq (t/N) {\sum}'
e^{i \vec{k} \cdot \vec{R}_{A} - i \vec{k}' \cdot \vec{R}_{B}}
\propto \delta(\vec{k} - \vec{k}')$.
For the uniform field (perpendicular to the graphene sheet)
of strength $10$~T, one can find
$a/(l^{2}_{B}|K-K^{\prime}|)\sim 10^{-4}$ so that the condition
(\ref{NOMIX}) is achieved, which is why the intervalley scattering
can be ignored in current experimental studies.
The intervalley mixing becomes important in a very strong magnetic
field $\sim 10^4$~T, where
$a/(l^{2}_{B}|K-K^{\prime}|)\sim 0.1$.

In the same way as above, one can find the condition when
the intervalley mixing is negligible in
a nonuniform field with constant gradient
$\lambda = \nabla B(\vec{r})$.
For the mixing to be ignored, the maximum value of the field
must satisfy the condition (\ref{NOMIX}).
In addition, the gradient $\lambda$ should be much smaller than
$10^4 \,\, \textrm{T} \cdot {\textrm{{\AA}}}^{-1}$, since
the path integration $(e/\hbar) \int_{\mathbf{C}} \vec{A} \cdot d \vec{r}_1$
in Eq. (\ref{TB_Hamiltonian})
becomes comparable to $\vec{K} \cdot \vec{R}_A - \vec{K}' \cdot \vec{R}_B$
when $\lambda = \lambda_0 \simeq 1.7 \times 10^4 \,\,
\textrm{T} \cdot {\textrm{{\AA}}}^{-1}$.
In current experimental studies, the gradient is much smaller than
$\lambda_0$ so that one can ignore the intervalley scattering.

\section{Transmission through a graphene ribbon interferometry}
\label{APP2}

In Appendix \ref{APP2}, we derive the transmission probability
through the inteferometry in Fig.~\ref{fig5}(a), based on the scattering
matrix formalism. The resulting expression in Eq. (\ref{tran})
can describe the Aharonov-Bohm effect of the interferometry.
One can easily obtain the transmission probability for other
setups with different edge channel configurations by slightly
modifying the derivation.

\begin{figure}
\includegraphics[width=0.45\textwidth]{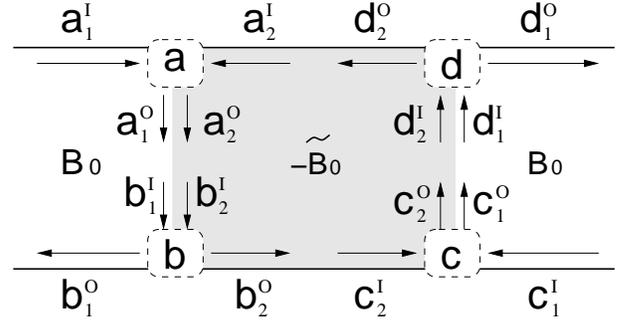}
\caption{
The same as Fig.~\ref{fig5}(a) but with detailed labeling
of edge channels (represented by solid arrows)
at its four scattering points $p \in \{ a,b,c,d \}$ (dashed boxes).
At each point $p$,
there are
two incoming channels with amplitude $(p_1^{\rm I}, p_2^{\rm I})$
and two outgoing ones with $(p_1^{\rm O}, p_2^{\rm O})$.
}\label{fig6}
\end{figure}

The interferometry is in the integer quantum Hall regime so
that its electron transport can be described by edge channels,
such as
the edge states along ribbon edges and
the magnetic edge states along the boundaries $\bar{ab}$ and $\bar{cd}$.
The scattering between the edge channels occurs at
four scattering points $p \in \{ a,b,c,d \}$.
Each point $p$ has
two incoming channels with amplitude $(p_1^{\rm I}, p_2^{\rm I})$
and two outgoing ones with $(p_1^{\rm O}, p_2^{\rm O})$;
for example,
the two incoming channels to the point $a$ are
one right-going and the other left-going channels along the upper
ribbon edge,
while the two outgoing channels from $a$
are the two magnetic edge channels along the line $\bar{ab}$
(see Fig.~\ref{fig6}).
At each point $p$,
we introduce
a scattering matrix $S_p$ which links the amplitudes
of the incoming and outgoing states,
\begin{equation}\label{sca}
\left(
\begin{array}{c}
 p^{\rm O}_{1} \\
 p^{\rm O}_{2} \\
\end{array}
\right)=
S_p
\left(
\begin{array}{c}
 p^{\rm I}_{1} \\
 p^{\rm I}_{2} \\
\end{array}
\right)=
\left(
\begin{array}{cc}
 s^{p}_{11} & s^{p}_{12} \\
 s^{p}_{21} & s^{p}_{22} \\
\end{array}
\right)
\left(
\begin{array}{c}
 p^{\rm I}_{1} \\
 p^{\rm I}_{2} \\
\end{array}
\right).
\end{equation}
The scattering matrix $S_{p}$ has a general form
of $2 \times 2$ unitary matrix,
\begin{equation}\label{psca}
\left(
\begin{array}{cc}
 s^{p}_{11} & s^{p}_{12} \\
 s^{p}_{21} & s^{p}_{22} \\
\end{array}
\right)=e^{i\theta_{p}}\left(
\begin{array}{cc}
 i\sqrt{1-t^{2}_{p}}e^{i\phi_{p}} & t_{p}e^{i\phi'_{p}} \\
 t_{p}e^{-i\phi'_{p}} & i\sqrt{1-t^{2}_{p}}e^{-i\phi_{p}} \\
\end{array}
\right).
\end{equation}
On the other hand, while edge channels propagate from one scattering
point $p$ to its neighboring point $p'$, they acquire phase accumulation
$\varphi_{pp'}$. As a result, one has
\begin{eqnarray}
(b_1^{\rm I}, b_2^{\rm I}) & = & e^{i \varphi_{ba}}
(a_1^{\rm O}, a_2^{\rm O}), \\
(d_1^{\rm I}, d_2^{\rm I}) & = & e^{i \varphi_{dc}}
(c_1^{\rm O}, c_2^{\rm O}), \\
 c^{\rm I}_{2}&=&e^{i\varphi_{cb}}b^{\rm O}_{2}, \\
  a^{\rm I}_{2}&=&e^{i\varphi_{ad}}d^{\rm O}_{2}. \label{PHA}
\end{eqnarray}
By combining the relations (\ref{sca})-(\ref{PHA})
and by putting $a_1^{\rm I} = 1$ and $c_1^{\rm I} = 0$,
one can obtain
the transmission probability $T = |d_1^{\rm O}|^2$ of
the edge state incoming from the source (the left of the inteferometry)
to the drain (the right),
\begin{equation}\label{tran}
T=\left|\frac{(t_b r_a + t_a r_b e^{-i\alpha})
(t_c r_d + t_d r_c e^{-i\beta})}
{1-e^{i\varphi}(t_a t_b - r_a r_b e^{-i\alpha})
(t_c t_d - r_c r_d  e^{-i\beta})}\right|^{2},
\end{equation}
where $r_p = \sqrt{1 - t_p^2}$,
$\varphi$ contains the Aharonov-Bohm phase as well as
the dynamical phase accumulated along one circulation of the closed loop $abcd$,
\begin{eqnarray}
\varphi&=&\varphi_{ba}+\varphi_{cb}+\varphi_{dc}+\varphi_{ad}
+\phi'_{a}-\phi'_{b}  +\phi'_{c}-\phi'_{d} \nonumber \\
& + & \sum_{p \in \{a,b,c,d \}} \theta_p,
\end{eqnarray}
$ \alpha = \phi_{a}+\phi_{b}+\phi'_{a}-\phi'_{b}$,
and
$\beta = \phi_{c}+\phi_{d}+\phi'_{c}-\phi'_{d}$.
The transmission $T$ can describe the Aharonov-Bohm oscillation
of the Dirac fermions in the setup of Fig.~\ref{fig5}(a),
as a function of $\tilde{B}$.
Note that in general the scattering matrix $S_p$ contains the information
of the scattering between the $K$ and $K'$ valleys.\cite{two}

We close this appendix by analyzing Eq. (\ref{tran}) for a simple case
of $t_a = t_b = t_c = t_d = t = 1 / \sqrt{2}$.
In this case, the transmission probability can be simplified as
\begin{equation}
T = \frac{\cos^2 \frac{\alpha}{2} \cos^2 \frac{\beta}{2}}{
1 + 2 \sin \frac{\alpha}{2} \sin \frac{\beta}{2} \cos (\varphi -
\frac{\alpha}{2} - \frac{\beta}{2}) + \sin^2 \frac{\alpha}{2} \sin^2 \frac{\beta}{2}}.
\end{equation}
This result shows a usual form for the Aharonov-Bohm
interference, except for the factors $\cos \alpha/2 \cos \beta /2$
and $\sin \alpha/2 \sin \beta /2$. From the factors, one can see that
there appears no interference whenever $\alpha = \pi$ or $\beta = \pi$.
It happens when
destructive interference occurs
during the propagation from one ribbon edge
to the other through the two magnetic edge channels
along $\bar{ab}$ or $\bar{cd}$
in such special cases.


\begin{thebibliography}{99}
\bibitem{EFFECTIVE} In the presence of the step potential,
$V^j_{\rm eff}$ depends on the
eigenenergy $E$ of the Dirac Hamiltonian
in Eq. (\ref{DIRAC});
for example, one finds $V_{\rm eff}^{j = A} = (x + k)^2/2
+ 1/2 + 2 E V_0 - V_0^2$ for $x < 0$.
As a result, $V_{\rm eff}^j$ should be studied in a self-consistent
way [see Eq. (\ref{DIRAC_SUB})].
\end{thebibliography}
\end{document}